\documentclass[prb,superscriptaddress,twocolumn,showpacs,preprintnumbers]{revtex4}
\usepackage{graphicx}
\usepackage{dcolumn}
\usepackage{bm}
\usepackage{braket}

\usepackage{color}

\begin{document}

\preprint{}
\title{Density matrix renormalization group study of optical conductivity in the 
one-dimensional Mott insulator Sr$_2$CuO$_3$}
\author{Shigetoshi Sota}
\author{Takami Tohyama}
\affiliation{Yukawa Institute for Theoretical Physics, Kyoto University, 606-8502, kyoto, Japan}
\date{\today}

\begin{abstract}
Applying newly developed dynamical density matrix renormalization group techniques at zero and
finite temperatures to a Hubbard-Holstein
model at half-filling, we examine the optical conductivity of a typical
one-dimensional Mott insulator Sr$_2$CuO$_3$.
We find a set of parameters in the Hubbard-Holstein model, which can describe optical conductivity for both Mott-gap
excitation in the high-energy region and phonon-assisted spin excitation in the low-energy region. We also find that electron-phonon interaction gives additional broadening in the temperature dependence of the Mott-gap
excitation.
\end{abstract}

\maketitle

\section{Introduction}
One-dimensional (1D) spin-1/2 quantum spin materials have attracted great attention
because they provide typical quantum spin systems with large quantum
fluctuations. A copper oxide compound Sr$_{2}$CuO$_{3}$ consists of
CuO chains with the $3d^{9}$ configuration of Cu ion and the $2p^6$ configuration of O ion, and it is known as a typical 
1D spin-1/2 Heisenberg system with only nearest-neighbor
interaction.~\cite{Maekawa} 
Antiferromagnetic exchange interaction $J$ between neighboring localized spins
is very large with $J\sim $0.26~eV (Refs.~2 and 3) 
but interchain exchange coupling is very weak. As a
result, three-dimensional long-range antiferromagnetic order is absent
down to very low temperature $T\sim $5~K, being regarded as an ideal 1D
spin-1/2 Heisenberg system.

Sr$_{2}$CuO$_{3}$ is a Mott insulator of charge-transfer (CT) type. 
The CT excitation appears at 1.8~eV in optical absorption spectra.~\cite{Ono} 
The CT excitation is known to be described by a single-band Hubbard
model with large on-site and nearest-neighbor Coulomb interactions, $U$ and $V
$,~\cite{Neudert,Matsueda} via mapping the Zhang-Rice singlet band onto the lower Hubbard band in the Hubbard model.~\cite{Maekawa} The CT excitation gap is, thus, called Mott gap. 
The optical conductivity above the CT excitation gap has been studied by using analytical and numerical methods.~\cite{Jeckelmann0, Essler, Controzzi}
The Hubbard model, however, cannot explain mid-infrared
absorption inside the Mott gap, which is originated from phonon-assisted spin
excitation.~\cite{Suzuura96, Lorenzana} In order to understand both the
Mott gap and spin excitations on the same footing, we need to treat the
Hubbard model with optical phonons. The simplest model is a
Hubbard-Holstein model containing Holstein-type coupling of electron to the
Einstein phonons.

In this paper, we examine the optical conductivity of the Hubbard-Holstein model by using a dynamical density-matrix renormalization group (DMRG) method~\cite{Jeckelmann} combined with a kernel polynomial expansion. By treating phonon degree of freedom as a quantum object, we calculate the optical conductivity of the model and reproduce both the Mott-gap  excitation and phonon-assisted spin excitation observed experimentally. We find a parameter set describing Sr$_{2}$CuO$_{3}$. Furthermore, by using a low-temperature dynamical DMRG method,~\cite{LTDMRG} we examine the temperature dependence of the Mott-gap excitation to clarify the effect of optical phonons on spectral shape at finite temperature. We find that the presence of phonons induces the enhancement of the width of an excitonic peak in the optical conductivity.

This paper is organized as follows. The Hubbard-Holstein model in 1D is introduced in Sec.~II. In Sec.~III, a dynamical DMRG combined with a kernel polynomial expansion is explained. The optical conductivity of the Hubbard-Holstein model at both zero and finite temperatures are shown in Sec.~IV. Summary is given in Sec.~V.

\section{Hubbard-Holstein model}
A coupling between an electron and breathing phonons in cuprates can be mapped onto Holstein- and Peierls-type electron-phonon (EP) interactions.~\cite{Gunnarsson} Since the Holstein interaction is stronger than the Peierls one, we consider a 1D Hubbard-Holstein model in the present work. In addition, we introduce a nearest-neighbor Coulomb repulsion that leads to excitonic effects. The Hubbard-Holstein Hamiltonian is defined by 
\begin{eqnarray}
H &=& -t\sum_{i,\sigma}(c_{i,\sigma}^{\dagger}c_{i+1,\sigma}+\mathrm{H.c.})
 + U\sum_{i}n_{i,\uparrow}n_{i,\downarrow} \nonumber \\
&& + V\sum_{i}(n_{i}-1)(n_{i+1}-1) \nonumber \\
&& + \omega_{0}\sum_{i}b_{i}^{\dagger}b_{i} -g\sum_{i}(b_{i}^{\dagger}+b_{i})(n_{i}-1),  
\label{H}
\end{eqnarray}
where $c_{i,\sigma}^{\dagger}$ ($c_{i,\sigma}$) is the creation (annihilation) operator of an electron at site $i$ with spin $\sigma$, and $b_{i}^{\dagger}$ ($b_{i}$) is the creation (annihilation) operator of a phonon at site $i$. This model includes electron hopping, $t$, on-site and nearest-neighbor
Coulomb repulsions, $U$ and $V$, respectively, phonon frequency, $\omega_{0}$, and EP coupling, $g$. Since the energy scale of the optical conductivity is much larger than the dispersion relation of the phonon, we neglect the dispersion of the phonon in order to make our discussion simple. Keeping cuprates in mind, we take $U$ larger than the band width $4t$. 

Here, we briefly mention how to construct the model Hamiltonian (\ref{H}) in the case of the cuprates. The lower Hubbard band of Eq.~(\ref{H}) corresponds to the Zhang-Rice singlet which is derived from a three-band model for Cu $3d_{x^{2}-y^{2}}$ and O $2p$ orbitals. Lattice vibration leads to changes in the hopping integral of an electron between neighboring Cu and O orbitals. The modulation of the hopping integral gives rise to a diagonal EP term, i.e., Holstein  term, in the effective single-band Hamiltonian, which is larger than an off-diagonal (Peierls) term.~\cite{Gunnarsson,Khaliullin}

In Hamiltonian (\ref{H}), we have introduced the $V$-term, $V(n_{i}-1)(n_{i+1}-1)$, instead of standard notation, $Vn_{i}n_{i+1}$. The latter term gives rise to an energy reduction by $-2V$, instead of $-V$, when a doubly occupied state (doublon) is located at the edge of a chain accompanied by a neighboring unoccupied state (holon). This induces a localization of the doublon-holon pair at the edge. On the other hand, for the former term, the energy reduction is always given by $-V$ for a doublon-holon pair irrespective of the location of the pair. Therefore, the $V(n_{i}-1)(n_{i+1}-1)$ term prevents the localization of the doublon-holon pair at the edge of a chain, leading to the reduction in a boundary effect.

\section{Dynamical DMRG combined with kernel polynomial expansion}
We examine the optical conductivity of the 1D Hubbard-Holstein model [Eq.~(\ref{H})] at half filling. Dynamical current-current correlation function reads
\begin{eqnarray}
\chi_j(\omega)&=&\frac{1}{\pi N_\mathrm{s} Z} \sum_{n} e^{-\beta \epsilon_n}
 \mathrm{Im} \bra{n}j^{\dagger}  \frac{1}{\omega - H - \epsilon_n - i\gamma}j\ket{n},\nonumber\\&&
\label{chi}
\end{eqnarray}
where $N_s$ is the number of electron sites, $Z$ is the partition function, $\beta$ is the inverse temperature, $j\equiv it \sum_{i, \sigma} (c_{i+1, \sigma}^\dagger c_{i, \sigma}-\mathrm{H.c.})$ is the current operator, $\ket{n}$ is an eigenstate with eigenvalue $\epsilon_n$, and $\gamma$ is a infinitesimally small energy. The optical conductivity is given by $\chi_j(\omega)/\omega$. Equation~(\ref{chi}) for the Holstein-Hubbard model is calculated by using a dynamical DMRG method~\cite{Jeckelmann} combined with a kernel polynomial expansion (shown below) at zero temperature and a low-temperature dynamical DMRG method~\cite{LTDMRG} at finite temperatures. 

The dynamical DMRG method employs a multitarget procedure. At zero temperature, the multitarget states corresponding to the optical conductivity [Eq.~(\ref{chi})] are $\ket{0}$, $j\ket{0}$, and the correction vector $\left[ \omega-H-\epsilon_0-i\gamma \right]^{-1} j\ket{0}$, where $\ket{0}$ represents the ground state.

There are several techniques to calculate the correction vector. Usually we use a conjugate gradient method extended to non-Hermitian matrix. In this case, the delta function is broaden by Lorentzian with the half width at half maximum (HWHM) of $\gamma$. Since $\gamma$ is usually taken to be small but finite number such as $\gamma=0.2t$ in our calculation at finite temperature, the tail of the Lorentzian extends to a wide range of energy. This tail sometimes covers a small spectral weight. Actually, the Mott-gap excitations at high energy tend to hide the phonon-assisted spin excitations at low energy, since the weight of the former is more than thousand times larger than that of the latter.~\cite{Tohyama} In order to diminish such a undesirable tail, we introduce a Gaussian broadening instead of the Lorentzian broadening.  

The introduction of the Gaussian broadening is achieved by employing a kernel polynomial expansion.~\cite{KPM} Using the Legendre polynomial $P_l(x)$, we expand the correction vector as
\begin{eqnarray}
&&\frac{1}{\omega-H-\epsilon_n-i\gamma}j\ket{n} \nonumber\\
&=&\int_{-\infty}^{-\infty}dx\frac{1}{\omega-x-\epsilon_n-i\gamma}\delta(x-H) j\ket{n} \nonumber\\
&=&\sum_{l=0}^{\infty} w_l^{-1} \int_{-1}^{1} dx \frac{ P_l(x)}{\omega^{\prime}-x-\epsilon_n^{\prime}-i\gamma}P_l(H_s) j\ket{n},
\label{cvo}
\end{eqnarray}
where $w_l=2/(2l+1)$ is a normalization constant for the orthogonality of the Legendre polynomial, $\omega^{\prime}=w_H \omega$, $\epsilon_n ^{\prime}=w_H \epsilon_n $, and $H_s=w_H H$, $w_H$ being a rescaling parameter to confine eigenvalues $\epsilon_n$ within the interval of $[-1,1]$. Note that the Legendre polynomial is defined on the real axis within this interval.

The integration in terms of $x$ in Eq.~(\ref{cvo}) reads
\begin{eqnarray}
\int_{-1}^1 dx \frac{P_l(x)}{\omega^{\prime}-x-\epsilon_n^{\prime}-i\gamma}
=2Q_l(\omega^{\prime}-\epsilon_n^{\prime}) + i\pi P_l(\omega^{\prime}-\epsilon_n^{\prime}),
\label{coef}
\end{eqnarray}
where $Q_l(\omega)$ is the Legendre polynomials of the second kind. From Eqs.~(\ref{cvo}) and (\ref{coef}), we may calculate the correction vector, provided that the Legendre polynomials are obtained by a three-term recursive formula. However, it is practically impossible to perform integration of $l$ in Eq.~(\ref{cvo}) up to infinity. Therefore, we need to introduce a truncation number $L$, whose value is usually several hundred.

The truncation of $l$ gives rise to so-called Gibbs oscillations in numerical data, which are known to be unphysical phenomena. Some improved kernel polynomials are suggested to remove the Gibbs oscillations.~\cite{KPM, Iitaka} In the present study, we employ a regulated polynomial~\cite{rpe} given by a Gauss distribution function, \begin{eqnarray}
\left<P_l(H_s)\right>_{\tilde{\sigma}}\equiv \frac{1}{\sqrt{2\pi \tilde{\sigma}^2}} \int_{-1}^{1} dx e^{-\frac{(x-H_s)^2}{2\tilde{\sigma}^2}}P_l(x),
\label{rp}
\end{eqnarray}
where $\tilde{\sigma}$ is the HWHM of the Gaussian given by $\tilde{\sigma}=2\pi/L$. The regulated polynomial can be recursively calculated by a three-term recursive formula without direct calculation of the Gaussian integral in Eq.~(\ref{rp}),
\begin{eqnarray}
\left<P_{l+1}(H_s)\right>_{\tilde{\sigma}}j\ket{n}&=&\frac{2l+1}{l+1}H_s \left<P_ln(H_s)\right>_{\tilde{\sigma}}j\ket{n}
\nonumber\\
&-&\frac{l}{l+1}\left<P_{l-1}(H_s)\right >_{\tilde{\sigma}}j\ket{n}
\nonumber\\
&+&\frac{2l+1}{l+1}\tilde{\sigma}^2\left<P_l^{\prime}(H_s)\right>_{\tilde{\sigma}}j\ket{n} \nonumber\\
\label{r1}
\end{eqnarray}
and
\begin{eqnarray}
\left<P_{l+1}^{\prime}(H_s)\right>_{\tilde{\sigma}}j\ket{n}&=&(2l+1)\left<P_l(H_s)\right>_{\tilde{\sigma}}j\ket{n}
\nonumber\\
&&+\left<P_{l-1}^{\prime}(H_s)\right>_{\tilde{\sigma}}j\ket{n},
\label{r2}
\end{eqnarray}
where $P_{l}^{\prime}(x)=dP_{l}(x) /dx$.
We note that this regulated polynomial method gives a Gaussian broadening of spectral weight in optical conductivity with HWHM of $\sigma=\tilde{\sigma}/\omega_H$ instead of the Lorentzian broadening with HWHM of $\gamma$ in the standard correction vector method.

\section{Results}
First, we determine a set of parameters of the Hubbard-Holstein model [Eq.~(\ref{H})] that reproduces well both phonon-assisted spin excitation and Mott-gap excitation simultaneously in the optical absorption of Sr$_2$CuO$_3$. 
The phonon energy $\omega_0$ is taken to be $\omega_0=0.11$~eV from experimental phonon peak at the bottom of phonon-assisted spin excitation.~\cite{Suzuura96} 
Since an excitonic peak seems to exist,~\cite{Neudert} we take a condition for $V$ to generate the excitonic peak in 1D Mott insulator: $V/t=2$.~\cite{Matsueda,Stephan} 
Remaining parameters are $t$, $U$, and $g$. From a comparison of optical conductivity between experiment and dynamical DMRG calculation, a relation of $U/t\sim 8$ was suggested.~\cite{Kim,Benthien} For $g$, a diagrammatic Monte Carlo simulation reported a good description of angle-resolved photoemission spectra in two-dimensional cuprates at half-filling when $g/t\sim0.4$,~\cite{Mishchenko} though a proper $g$ value may depend on the value of $\omega_0$. Turning the ratios of $U/t$ and $g/t$ together with the value of $t$, we find a best parameter set that can describe both the Mott-gap and phonon-assisted spin excitations in different energy regions. The best parameter set obtained is $t=0.41$~eV, $U=3.3$~eV, $V=0.82$~eV, $g=0.16$~eV, and $\omega_0=0.11$~eV. The exchange interaction given by a $1/U$ expansion, $J=4t^2/(U-V)=0.273$~eV, is close to the experimentally estimated values, $J\sim 0.26$~eV.~\cite{Motoyama96,Suzuura96} We note that the parameters of $t$, $U$, and $V$ are different from those in Ref.~17, 
although the difference is not significant.

\begin{figure}[tb]
\includegraphics[scale=0.45]{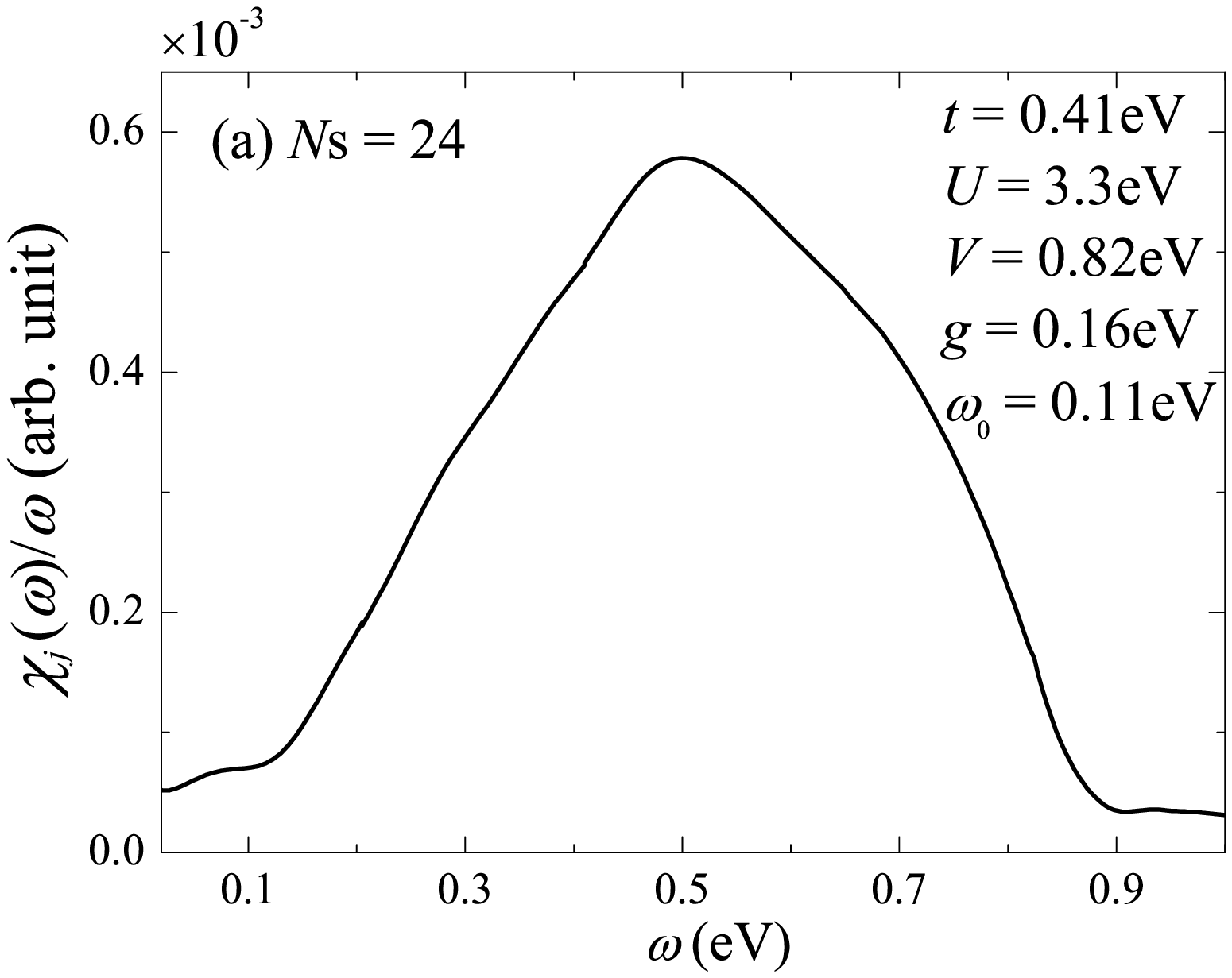}
\includegraphics[scale=0.45]{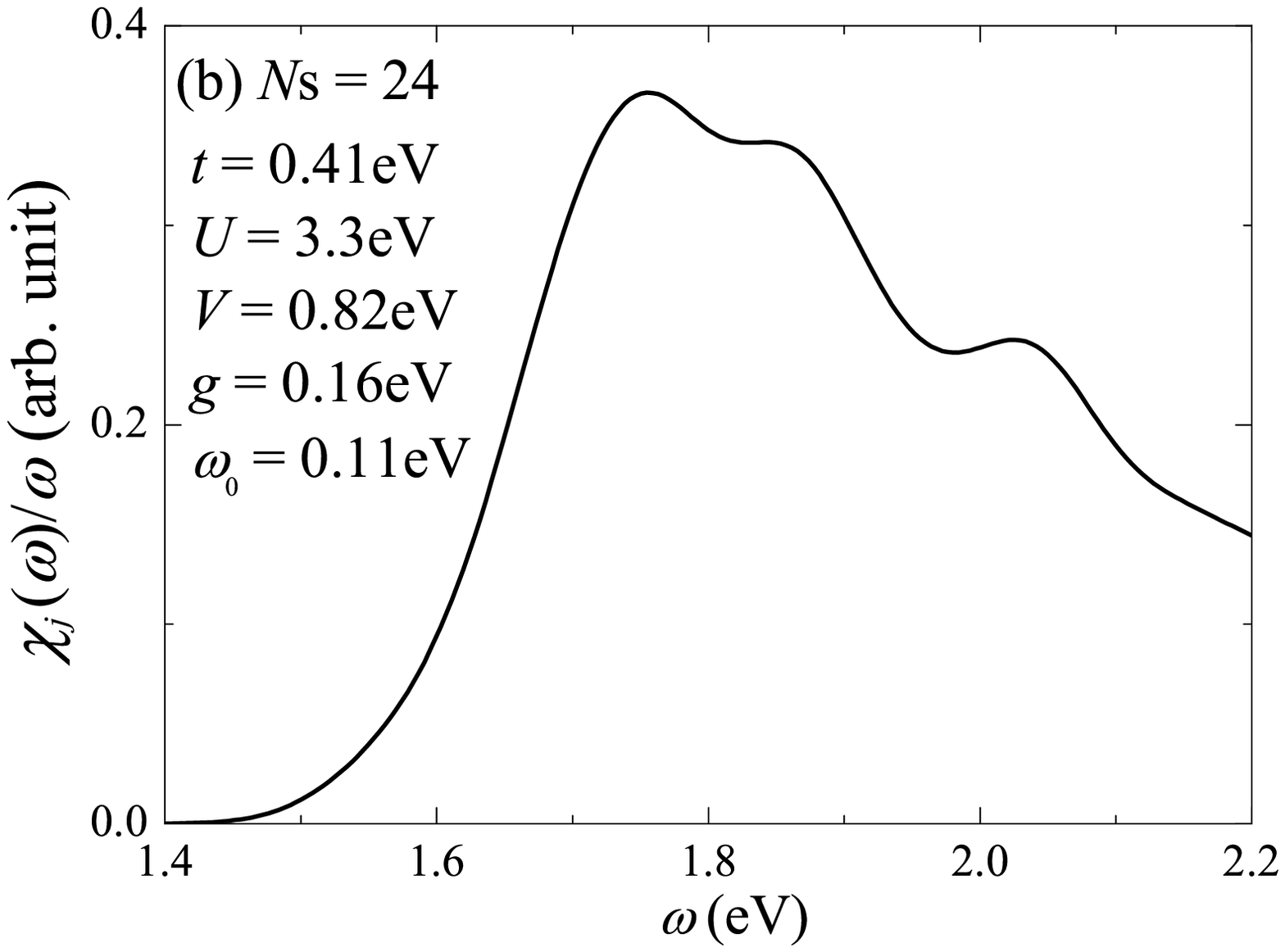}
\caption{\label{fop} Optical conductivity in a Hubbard-Holstein chain with 24 sites. (a) Phonon-assisted spin excitation, and (b) Mott-gap excitation. The broadening factor $\sigma=0.04$~eV. The DMRG truncation number $m=600$ and $m=800$ for (a) and (b), respectively. Parameter values are shown in the panels. 
}
\end{figure}

Figure~\ref{fop} shows optical conductivity at zero temperature for a 24-site chain under open boundary condition. We note that similar spectral behaviors in the optical conductivity are obtained for a smaller 20-site chain (not shown). The Gaussian broadening  $\sigma$ is taken to be $\sigma=0.04$~eV, whose value is enough to smear out discrete weights due to finite-size effect [see the inset of Fig.~\ref{fop}(b)].
The number of the states kept in the DMRG procedure ($m$, the DMRG truncation number) is set to be $m=600$ and $800$, which is enough to get good convergency.

Figure~\ref{fop}~(a) shows phonon-assisted spin excitations in the optical conductivity. 
A phonon peak appears at $\omega_0=0.11$~eV as expected. Just above the phonon peak, a broad structure emerges as phonon-assisted spin excitation. Without EP coupling, we cannot obtain this structure. The energy position of a broad peak, $\omega \sim 0.48$~eV, is consistent with an experimental value.~\cite{Suzuura96}  The peak comes from a Van Hove singularity of spinon excitation.~\cite{Suzuura96,Lorenzana2}

Figure~\ref{fop}~(b) shows the optical conductivity in the CT energy region. A peak appears at $\omega=1.75$~eV. This energy is in agreement with the experimental data.~\cite{Ono} In addition, we can find a hump structure at $\omega=1.86$~eV. It is natural to assign the hump structure to a one-phonon excitation on top of the $\omega=1.75$~eV structure because of $\omega_0=0.11$~eV. Thus EP interaction contributes to the broadening of the main peak in the optical conductivity.~\cite{Matsueda2}

The intensity of the Mott-gap excitation is several hundred times larger than that of the phonon-assisted spin excitation, which is consistent with the experimental data.~\cite{Suzuura96} Judging from the agreement of calculated spectral weights with experimental ones for both the Mott-gap excitation and phonon-assisted spin excitation, we conclude that the Hubbard-Holstein model with the suggested parameter set can describe very well the optical properties of Sr$_2$CuO$_3$.

\begin{figure}[tb]
\includegraphics[scale=0.45]{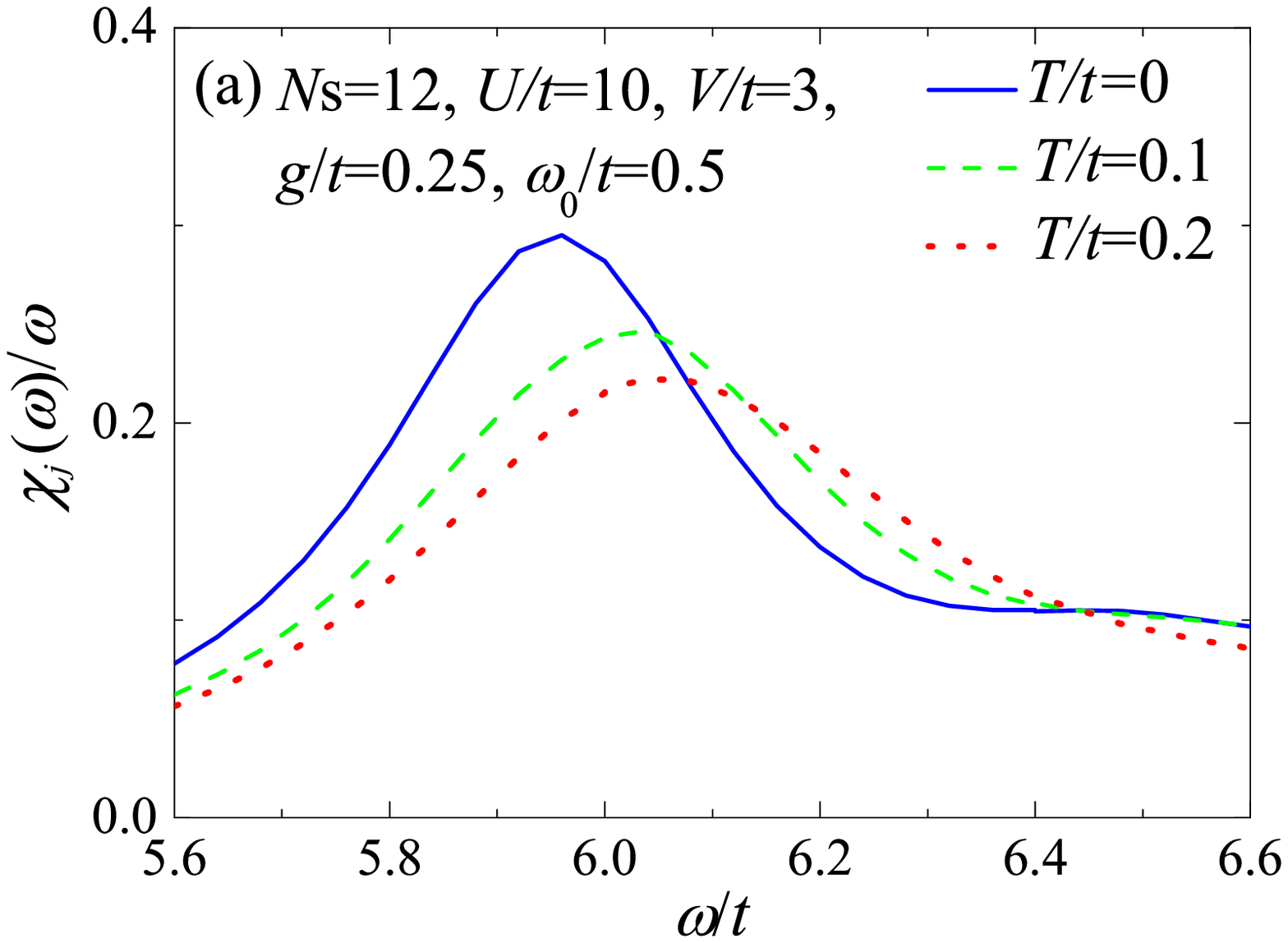}
\includegraphics[scale=0.43]{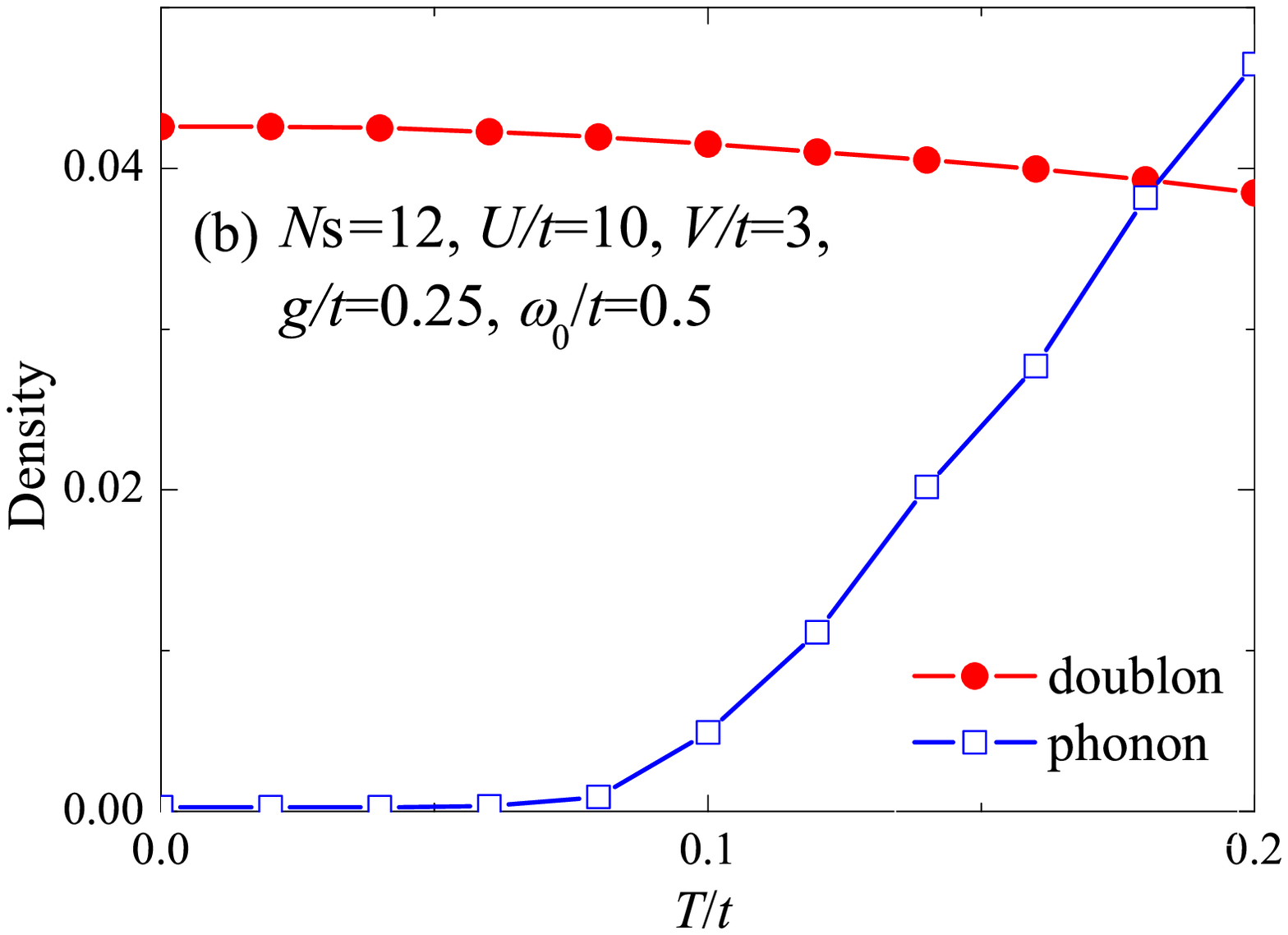}
\caption{\label{ftd} (Color online) Temperature dependence of (a) the optical conductivity and (b) the density of doublon and phonon in a 12-site Hubbard-Holstein chain. The truncation number $m=400$. A Lorentzian broadening factor $\gamma=0.2t$ in (a).}
\end{figure}

Next we examine the effect of temperature on the optical conductivity in 1D Mott insulators. Figure~\ref{ftd} shows the temperature dependence of the optical conductivity in the Mott-gap energy region for system size $N_s=12$. 
Since the Herbert space of the $N_s=12$ system is very large due to phonon degrees of freedom, it is impossible to perform fully exact diagonalization even though the system size is small. Then we use the low-temperature dynamical DMRG procedure. 
In this calculation, we use the conjugate gradient method to obtain the correction vector with a broadening factor of $\gamma=0.2t$ in Eq.~(\ref{chi}). 
We employ a parameter set of the Hubbard-Holstein model as $U/t=10$, $V/t=3$, $\omega_0/t=0.5$, and $g/t=0.25$, instead of the best parameter set for Sr$_2$CuO$_3$: $U/t=8$, $V/t=2$, $\omega_0/t=0.27$, and $g/t=0.39$. We have chosen this new parameter set to reduce computational costs because a larger $\omega_0/t$ as well as a smaller $g/t$ reduces the number of phonons to be included in our numerical calculations. The values of parameters are, however, close to the values estimated for Sr$_2$CuO$_3$. Thus, it is expected that the tendency of the temperature dependence of the optical conductivity is similar to that obtained by the best parameter set for Sr$_2$CuO$_3$ used in Fig.~1. 
The truncation number is $m=400$. The solid line in Fig.~\ref{ftd}~(a) shows the result at zero temperature. Since $V/t=3$ is larger than $V/t=2$ in Fig.~~\ref{fop}~(b), the excitonic peak at $\omega/t=5.95$ is more isolated from high-energy spectral weight consisting of phonon-related states as well as continuous states of the Mott-gap excitation. The broken and the dotted lines in Fig.~\ref{ftd}~(a) represent the results at $T=0.1t$ and $0.2t$, respectively. We find that the intensity of the peak decreases with increasing temperature, which is consistent with the results obtained experimentally.~\cite{Ono} 
It is noted that the peak position shifts toward higher $\omega$ with increasing temperature. The shift comes from the finite size effect, which was confirmed by investigating the system size dependence.~\cite{Onodera}

\begin{figure}[tb]
\includegraphics[scale=0.45]{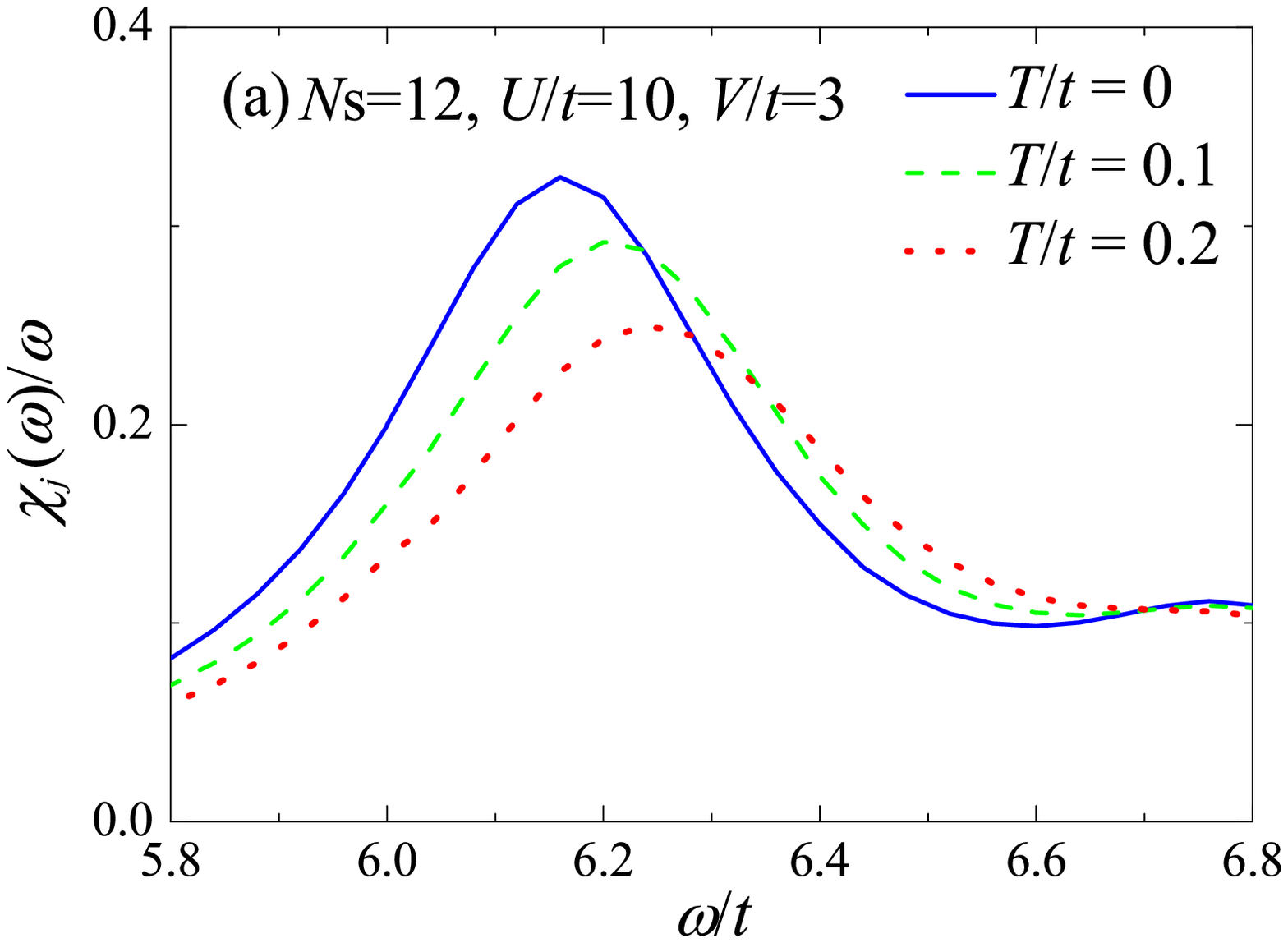}
\includegraphics[scale=0.45]{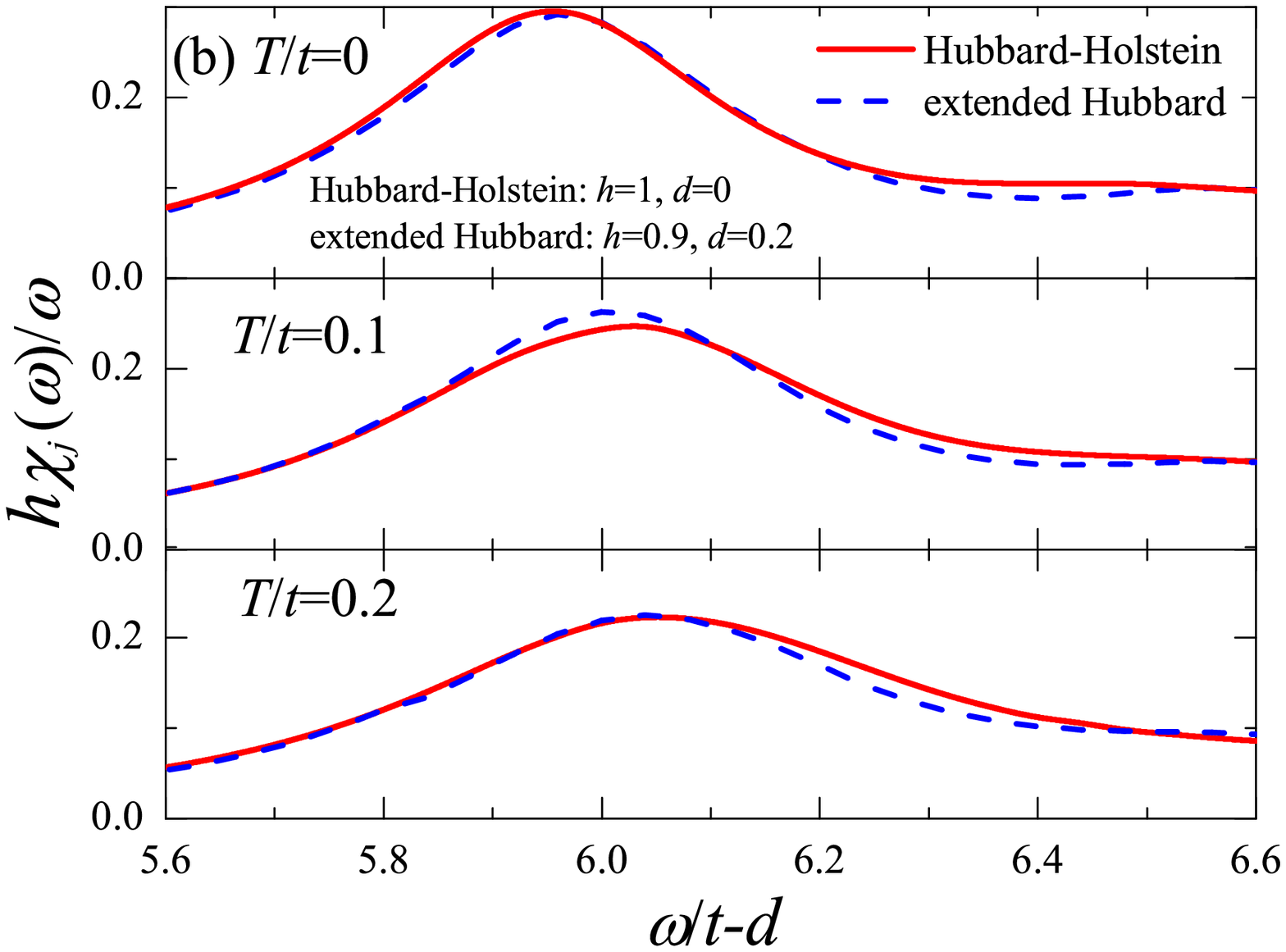}
\caption{\label{fpd} (Color online)  (a) The temperature dependence of the optical conductivity in a 12-site  extended Hubbard chain. (b) Comparison between the temperature dependence of the optical conductivity in the Hubbard-Holstein model (solid lines) and that in the extended Hubbard model (dashed lines).} 
\end{figure}

To investigate the detail of such a temperature effect, we calculate the number of phonon per site, $N_\mathrm{s}^{-1}\sum_n e^{-\beta\epsilon_n}\braket{n|\sum_{i}b_{i}^{\dagger}b_{i}|n}$ and double occupation of electron per site (the number of doublon), $N_\mathrm{s}^{-1}\sum_n e^{-\beta\epsilon_n}\braket{n|\sum_{i}c_{i,\uparrow }^{\dagger}c_{i,\uparrow }c_{i,\downarrow }^{\dagger}c_{i,\downarrow }|n}$. Figure~\ref{ftd}~(b) shows their temperature dependence up to $0.2t\sim J/2$. We find that the phonon number gradually increases with increasing temperature, while the doublon number only slightly changes. It is natural to consider that phonons excited by temperature give some effects on the spectral shape of the optical conductivity.

The intensity of the spectra in both Figs.~2(a) and 3(a) decreases with increasing temperature. The reason of such temperature dependence comes from the increase in the probability of ferromagnetic alignment of neighboring spins, which prevents charge transfer process due to the electric field. However, phonons may also have a role in the temperature dependence of the optical conductivity.  
In order to make clear the role of the phonons, we calculate the temperature dependence of the optical conductivity for the extended Hubbard model without the EP interaction in Eq.~(\ref{H}) and show the results in Fig.~\ref{fpd}~(a). 
Comparing Fig.~\ref{fpd}~(a) with Fig.~\ref{ftd}(a), one can find that the peak at finite temperatures in Fig.~\ref{ftd}(a) seems to be wider than that in Fig.~\ref{fpd}(a). To see this more clearly, we fit the peak position of the $T=0$ spectrum in the extended Hubbard model to that in the Hubbard-Holstein model. Here we introduce scaling parameters for energy shift, $d$, and for hight change, $h$. We obtain $d=0.2$ and $h=0.9$. The rescaled result is shown in Fig.~\ref{fpd}(b) as dashed lines. We find that spectral weight of the Hubbard-Holstein model at around $\omega/t-d=6.45$ ,
which is higher by $\omega_0$ than the peak position at $\omega =5.95$,
is larger than that of the extended Hubbard model.    
This is due to the presence of phonon structure as discussed in Fig.~\ref{fop}(b). 
The same scaling parameters are applied to the $T=0.1t$ and $T=0.2t$ spectra for the extended Hubbard model. We clearly find that the peak width of the Hubbard-Holstein model at $T=0.2t$ is wider than that of the extended Hubbard model. At $T=0.1t$, although the difference of the peak width between the Hubbard-Holstein model and the extended Hubbard model is smaller than that at $T=0.2t$, the tendency in the peak width is same as that at $T=0.2t$. This implies that the EP interaction broadens the peak structure 
with increasing temperature as a consequence of the enhancement of phonon density with $T$ as shown in Fig.~\ref{ftd}(b). From this result, we find that the EP interaction cannot be ignored when we discuss the temperature dependence of the optical conductivity in 1D Mott insulator Sr$_2$CuO$_3$.

\section{Summary}
We have investigated the optical conductivity in 1D Mott insulator Sr$_2$CuO$_3$. We have employed the Hubbard-Holstein chain including the Holstein-type EP interaction and Einstein phonon. Using a newly developed dynamical DMRG technique combined with a kernel polynomial expansion, we have found that, for a proper parameter set, our model reproduces simultaneously both Mott-gap excitation at the high-energy region and the phonon-assisted spin excitation at the low-energy region in the optical conductivity. We conclude that the Hubbard-Holstein model provides a good description of Sr$_2$CuO$_3$, and thus the EP interaction plays important roles in the electronic structure of Sr$_2$CuO$_3$. Using the low-temperature dynamical DMRG technique, we have found that the EP interaction broadens a peak structure in the optical conductivity with increasing temperature. This is accompanied by the increase in phonon number. Thus the EP interaction also plays an important role in the temperature dependence of the optical conductivity in the 1D Mott insulator such as Sr$_2$CuO$_3$.

\acknowledgements
The authors thank H. Matsueda for useful discussions. 
This work was supported by Next Generation Supercomputing Project of Nanoscience Program, Grant-in-Aid for Scientific Research (22340097) from MEXT, and the Academic Center for Computing and Media Studies, Kyoto University (ACCMS) for the use of the computing facilities. The numerical calculations were carried out at YITP and ACCMS, Kyoto University, and ISSP, University of Tokyo.


\begin{thebibliography}{99}
\bibitem{Maekawa}
See, for instance, S, Maekawa, T. Tohyama, S. E. Barnes, S. Ishihara, W. Koshibae, and G. Khaliullin, {\it Physics of Transition Metal Oxides} (Springer-Verlag, Berlin, 2004).

\bibitem{Motoyama96} N. Motoyama, H. Eisaki, and S. Uchida, Phys. Rev. Lett. {\bf 76}, 3212 (1996).

\bibitem{Suzuura96} H. Suzuura, H. Yasuhara, A. Furusaki, N. Nagaosa, and Y. Tokura, Phys. Rev. Lett. {\bf 76}, 2579 (1996).

\bibitem{Ono}
M. Ono, K. Miura, A. Maeda, H. Matsuzaki, H. Kishida, Y. Taguchi, Y. Tokura, M. Yamashita, and H. Okamoto, Phys. Rev. B {\bf 70}, 085101 (2004).

\bibitem{Neudert} R. Neudert, M. Knupfer, M. S. Golden, J. Fink, W. Stephan, K. Penc, N. Motoyama, H. Eisaki, and S. Uchida, Phys. Rev. Lett. {\bf 81}, 657 (1998).

\bibitem{Matsueda}
H. Matsueda, T. Tohyama, and S. Maekawa, Phys. Rev. B {\bf 70}, 033102 (2004); Phys. Rev. B {\bf 71}, 153106 (2005).
 
\bibitem{Jeckelmann0}E. Jeckelmann, F. Gebhard, and F. H. L. Essler, Phys. Rev. Lett. {\bf 85}, 3910 (2000).

\bibitem{Essler}F. H. L. Essler, F. Gebhard, and E. Jeckelmann, Phys. Rev. B. {\bf 64}, 125119 (2001).

\bibitem{Controzzi}D. Controzzi, F. H. L. Essler, and A. M. Tsvelik, Phys. Rev. Lett {\bf 86}, 680 (2001).

\bibitem{Lorenzana}J. Lorenzana and G. A. Sawatzky, Phys. Rev. Lett. {\bf 74}, 1867 (1995); Phys. Rev. B {\bf 52}, 9576 (1995).

\bibitem{Jeckelmann} E. Jeckelmann, Phys. Rev. B {\bf 66}, 045114 (2002).

\bibitem{LTDMRG} S. Sota and T. Tohyama,  Phys. Rev. B {\bf 78}, 113101 (2008).

\bibitem{Gunnarsson}
O. R\"{o}sch and O. Gunnarsson, Phys. Rev. Lett. {\bf 92}, 146403 (2004); Phys. Rev. B {\bf 70}, 224518 (2004).

\bibitem{Khaliullin} G. Khaliullin and P. Horsch, Physica C {\bf{282-287}}, 1751 (1997); P. Horsch and G. Khaliullin, Physica B {\bf{359-361}}, 620 (2005); P. Horsch, G. Khaliullin, and V. Oudovenko, Physica C {\bf{341-348}}, 117 (2000).

\bibitem{Tohyama}T. Tohyama and H. Matsueda, Prog. Theor. Phys. Suppl. {\bf 176}, 165 (2008).

\bibitem{KPM} A. Wei\ss e, G. Wellein, A. Alvermann, and H. Fehske, Rev. Mod. Phys. {\bf 78}, 275 (2006).

\bibitem{Iitaka} T. Iitaka and T. Ebisuzaki, Phys. Rev. Lett. {\bf 90}, 047203 (2003).

\bibitem{rpe} S. Sota and M. Itoh, J. Phys. Soc. Jpn. {\bf 76}, 054004 (2007).

\bibitem{Stephan} W. Stephan and K. Penc, Phys. Rev. B {\bf 54}, R17269 (1996).

\bibitem{Kim} Y.-J. Kim, J. P. Hill, H. Benthien, F. H. L. Essler, E. Jeckelmann, H. S. Choi, T. W. Noh, N. Motoyama, K. M. Kojima, S. Uchida, D. Casa, and T. Gog, Phys. Rev. Lett. {\bf 92}, 137402 (2004).

\bibitem{Benthien}H. Benthien and E. Jeckelmann, Phys. Rev. B {\bf 75}, 205128 (2007).

\bibitem{Mishchenko} A. S. Mishchenko and N. Nagaosa, Phys. Rev. Lett. {\bf 93}, 036402 (2004).

\bibitem{Lorenzana2}J. Lorenzana and R. Eder, Phys. Rev. B {\bf 55}, R3358 (1997).

\bibitem{Matsueda2}H. Matsueda, A. Ando, T. Tohyama, and S. Maekawa, Phy. Rev. B {\bf 77}, 193112 (2008).

\bibitem{Onodera}H. Onodera, T. Tohyama, and S. Maekawa, Phys. Rev. B {\bf 69}, 245117 (2004).

\end{thebibliography}
\end{document}